\documentclass[12pt]{article}

\usepackage{graphicx}
\begin{document}
\begin{center}
{\Large Acceleration of the Universe, String Theory and a Varying
Speed of Light }
\vskip 0.3 true in {\large J. W. Moffat} \date{} \vskip
0.3 true in {\it Department of Physics, University of Toronto, Toronto,
Ontario M5S 1A7, Canada}

\end{center}

\begin{abstract}%
The existence of future horizons in spacetime geometries poses serious
problems for string theory and quantum field theories. The observation
that the expansion of the universe is accelerating has recently been shown
to lead to a crisis for the mathematical formalism of string and
M-theories, since the existence of a future horizon for an eternally
accelerating universe does not allow the formulation of physical S-matrix
observables. Postulating that the speed of light varies in an
expanding universe in the future as well as in the past can eliminate
future horizons, allowing for a consistent definition of S-matrix
observables.

\end{abstract} \vskip
0.2 true in e-mail: moffat@medb.physics.utoronto.ca

\section{\bf Introduction}

Recently, it has been shown that a critical situation arises in string and
M-theories due to the existence of future horizons in
spacetimes~\cite{Susskind,Fischler,Witten}, and for models of the universe
exhibiting an accelerating expansion~\cite{Perlmutter}. String theories and
M-theories, in their present mathematical frameworks, are critically
dependent on their background geometries. The observables of string theory
are determined for asymptotically free particle states in asymptotically
flat Minkowski spacetime. The existence of an S-matrix is crucially
dependent on having a large enough space at infinity in which particles are
separated into a system of non-interacting objects. The description of
observables in AdS spacetimes by boundary correlators of bulk fields is
analogous to S-matrix elements, and is supported by an infinite asymptotic
space with non-interacting particles at infinity.

The standard quantum field theory formalism, which provides the basis for
the remarkable standard model agreement with observational data, is a
perturbative theory based largely on the notion of asymptotically free
particles and fields. In $dS$ spacetime with a positive cosmological
constant problems arise, for there is no unique Fock space vacuum and it is
difficult to define momentum space due to the ill-defined nature of
particle annihilation and creation operators. The indication that
observational data tell us that the universe is accelerating produces a
crisis for our conventional quantum field theory formalism. The
perturbative and non-perturbative string theories, formulated for strictly
on-shell S-matrix elements, do not fare any better and are possibly in
worse shape due to their on-shell definitions. This crisis in our
interpretation of modern particle theories has been waiting to happen; the
advent of observational data supporting an accelerated expansion of the
universe, forces us to confront particle theory with the complications of
spacetimes with future horizons.

Realistic cosmology models are described by neither flat
spacetime nor $AdS$ spacetime. Spatially flat FRW models can be described
by S-vectors as suggested by Witten~\cite{Witten}, by assuming that the
initial state of the universe is unique and that the final state is
described by asymptotically free particles with a Fock space of asymptotic
out-fields. In standard FRW models there is no future particle horizon, so
that particles can communicate with particles at infinity and an S-vector
can be meaningfully constructed.

It is possible that string theory can be completely reformulated, so that
it can cope with spacetimes with future horizons~\cite{Banks}. On the other
hand, it is possible that if the data supporting an accelerating universe
are confirmed, then we may not be able to reformulate the language of
quantum field theory and string theories in a satisfactory way, so that we
are forced to consider new ideas that can resolve the crisis. In the
following, we shall explore the idea that a varying light speed in the
{\it future} universe can remove future horizons from all spacetime
geometries and rid us of the challenging (and maybe impossible) problem
of making sense of present theories in cosmological backgrounds with
future horizons.

The idea that the speed of light varies was proposed as an alternative to
standard inflation theory
~\cite{Moffat,Moffat2,Albrecht,Barrow,Barrow2,Brandenberger,
Avelino,Hark,Magueijo,Clayton,Drummond,Liberati,Kiritsis}.
The idea~\cite{Moffat} originated with the hypothesis that there
is a phase of spontaneously broken, local Lorentz invariance in the early
universe, due to a non-vanishing vacuum expectation value of a field. The
speed of light underwent an abrupt phase transition as the universe
expanded decreasing to its presently observed value. This idea was
reformulated as a bimetric theory based on vector-tensor and scalar-tensor
structures~\cite{Clayton,Drummond,Liberati}. The notion of a varying speed
of light (VSL) has also been formulated in theories with extra-dimensions
and in 5-dimensional brane-bulk models with violations of Lorentz
invariance~\cite{Kiritsis}. There are observational indications that the
fine-structure constant $\alpha=e^2/\hbar c$ varies with time consistent
with an increasing speed of light~\cite{Webb}.

In the following, we shall show that VSL theories can remove the problem of
future horizons, allowing for a physical framework to define a consistent
S-matrix as a basis for quantum field theory and string/M-theories.

\section{\bf Varying Speed of Light Model}

We shall use a minimal scheme proposed in
refs.~\cite{Albrecht,Barrow,Barrow2} to illustrate the resolution of the
future horizon problem, and defer the application of a more geometrically
rigorous theory of VSL, such as the bimetric
theory~\cite{Clayton,Drummond,Liberati} to a later publication. In a
minimally coupled VSL theory, one replaces $c$ by a field in a preferred
frame of reference, $\chi(x^\mu)=c^4$. The dynamical variables in the
Lagrangian ${\cal L}$ are the metric $g_{\mu\nu}$, matter variables
contained in the matter Lagrangian ${\cal L}_M$, and the scalar field
$\chi$ which is assumed not to couple to the metric explicitly. In the
preferred frame the curvature tensor is to be calculated from $g_{\mu\nu}$
at constant $\chi$ in the normal manner. Varying the action with respect to
the metric gives the field equations
\begin{equation}
G_{\mu\nu}-g_{\mu\nu}\Lambda=\frac{8\pi G}{\chi}T_{\mu\nu},
\end{equation}
where $G_{\mu\nu}=R_{\mu\nu}-\frac{1}{2}g_{\mu\nu}R$, $\Lambda$ is the
cosmologcal constant and $T_{\mu\nu}$ denotes the stress energy-momentum
tensor. This theory is not locally Lorentz invariant. Choosing a specific
time to be the comoving proper time, and assuming that the universe is
spatially homogeneous and isotropic, so that $c$ only depends on time
$c=c(t)$, then the FRW metric can still be written as
\begin{equation}
ds^2=c^2dt^2-a^2\biggl(\frac{dr^2}{1-kr^2}+r^2d\Omega^2\biggr),
\end{equation}
where $k=0,+1,-1$ for spatially flat, closed and open
universes, respectively. The Einstein equations are still of the form
\begin{equation}
\biggl(\frac{{\dot a}}{a}\biggr)^2=\frac{8\pi
G}{3}\rho-\frac{kc^2}{a^2},
\end{equation}
\begin{equation}
\label{accelerationeq} \frac{{\ddot a}}{a}=-\frac{4\pi
G}{3}\biggl(\rho+3\frac{p}{c^2}\biggr).
\end{equation}
We have set
$\Lambda=0$, since we shall be primarily concerned with quintessence
models in which it is assumed that the cosmological constant is zero. The
conservation equations are modified due to the time dependence of $c$:
\begin{equation}
\label{conservationeq} {\dot\rho}+3\frac{{\dot
a}}{a}\biggl(\rho+\frac{p}{c^2}\biggr) =\frac{3kc^2}{4\pi Ga^2} \frac{{\dot
c}}{c}.
\end{equation}

Let us assume that matter obeys the equation of state
\begin{equation}
p=w\rho c^2,
\end{equation}
where $w$ is a constant. We shall assume that $c$ changes at a rate
proportional to the expansion of the universe
\begin{equation}
c={\bar c}a^n,
\end{equation}
where ${\bar c}$ and $n$ are constants. The present observational value of
$c$ is defined to be
\begin{equation}
c(t_0)\equiv c_0={\bar c}a^n(t_0),
\end{equation}
where $t_0$ denotes the present time.
Barrow~\cite{Barrow} has found an exact
solution of (\ref{conservationeq}) of the form
\begin{equation}
\rho=\frac{B}{a^{3(1+w)}}+\frac{3k{\bar c}^2na^{2(n-1)}} {4\pi
G(2n-2+3(1+w))},
\end{equation}
where $B \geq 0$ is constant if
$2n-2+3(1+w)\not= 0$. For $n=0$ the speed of light is constant and the
equations reduce to the usual adiabatic expansion laws of FRW cosmology. As
first shown in ref.~\cite{Moffat} and subsequently in
refs.~\cite{Albrecht,Barrow,Clayton} VSL theories can solve the horizon,
flatness and particle relic problems of early universe cosmology.

\section{\bf Future Horizons, Quintessence and Varying Speed of Light}

We shall adopt the theory of quintessence~\cite{Peebles} as an
alternative to a positive cosmological constant. According to this
theory, the dark energy of the universe is dominated by the potential
$V(\phi)$ of a scalar field $\phi$, which rolls down to its minimum at
$V=0$. We recall that a cosmological constant corresponds to $w=-1$,
radiation domination to $w=\frac{1}{3}$ and matter domination to $w=0$. On
the other hand, quintessence gives an equation of state with
\begin{equation}
\label{wrange} -1 < w < -\frac{1}{3},
\end{equation} while
the observational evidence for a cosmological constant is given by the
bound:
\begin{equation}
-1 < w_{\rm observed} \leq -\frac{2}{3}.
\end{equation}
We shall now analyze the causal structure of the universe
with $w$ in the range (\ref{wrange}) with a varying light speed $c=c(t)$.
Our results can be straightforwardly extended to higher-dimensional
theories such as brane-bulk models.

We obtain from (\ref{accelerationeq}) the
condition for an accelerating expansion of the universe, $p< -\rho/3$. The
proper horizon distance is given by
\begin{equation}
\delta_H(t)=a(t)I,
\end{equation}
where
\begin{equation}
I=\int^\infty_{t_0}\frac{dt'c(t')}{a(t')}.
\end{equation}
Whenever $I$ diverges there exist no future event horizons
in the spacetime geometry. On the other hand, when $I$ converges, the
spacetime geometry exhibits a future horizon, and events whose coordinates
at time ${\bar t}$ are located beyond $\delta_H$ can never communicate with
the observer at $r=0$.

The variation of the expansion scale factor at large $a(t)$, when the
curvature becomes negligible, approaches
\begin{equation}
a(t)\sim t^{2/3(1+w)}.
\end{equation}
We now have
\begin{equation}
\label{Iintegral}
I={\bar c}\int^\infty_{t_0}dt^\prime t^{\prime[2(n-1)/3(1+w)]}.
\end{equation}
We see that for the quintessence $w$ range (\ref{wrange}), we can choose
$n$ so that $I$ diverges and the future horizon has been eliminated. For
$n=0$, $I$ will converge for the quintessence $w$ range and will generate a
future horizon, which prohibits an S-matrix description of particles.
Consider, as an example, the choice $w=-\frac{2}{3}$, then
(\ref{Iintegral}) diverges for $n\geq \frac{1}{2}$ and the future horizon
is eliminated.

Consider finally a $dS$ universe with $\Lambda > 0$ and the asymptotic
scale factor behavior
\begin{equation}
a(t)\sim\exp\biggl(\sqrt{\frac{\Lambda}{3}}t\biggr).
\end{equation}
We now get
\begin{equation}
I={\bar c}\int_{t_0}^\infty
dt^\prime\exp\biggl[(n-1)\sqrt{\frac{\Lambda}{3}}t^\prime\biggr]
\end{equation}
and the $dS$ event horizon is removed for $n\geq 1$.

\section{\bf Conclusions}

We have concerned ourselves in this note with the serious difficulty in
defining a consistent S-matrix description of quantum field theory and
String/M-theories in spaces
associated with an eternal accelerated expansion of the universe. We have
shown that if we postulate that the speed of light varies in the future as
well as in the past universe, then we can solve the initial value problems
of cosmology (horizon and flatness problems), and remove asymptotically all
future horizons associated with quintessence models and an accelerating
universe. If future data confirm the accelerated expansion of the universe,
and it proves impossible to formulate a consistent theory of quantum
fields, strings and M-theory, based on physically meaningful quantum
observables, then we may be forced to seriously consider a scenario such as
that provided by VSL theories to preserve our present understanding of
particle physics and future theories of quantum gravity.

Hopefully, new supernovae red-shift data will be able to distinguish
between quintessence models and a non-vanishing cosmological constant
$\Lambda$. Whatever the outcome of these observations, the VSL
theories can eliminate the problem of asymptotic states in string
theory and quantum field theory, whether or not quintessence models can be
found that lead in the future to a decelerating universe or whether the
cosmological constant is non-zero, resulting in an eternal accelerating
universe.

\vskip
0.3 true in

{\bf Acknowledgment}
\vskip 0.2 true in
This work was supported by the Natural Sciences and Engineering Research
Council of Canada.
\vskip 0.5 true in

\end{document}